\documentclass[a4paper,11pt]{article}
\usepackage{jheppub} 
\usepackage[T1]{fontenc} 
\usepackage{epsfig}
\usepackage{subcaption}
\usepackage{color}
\usepackage{float}
\usepackage{slashed}
\usepackage{accents}
\def\beq{\begin{equation}}
\def\eeq{\end{equation}}
\def\beqn{\begin{eqnarray}}
\def\eeqn{\end{eqnarray}}

\def\met{E_T \hspace*{-1.1em}/\hspace*{0.5em}}

\newcommand*{\pbar}[1]{\accentset{(-)}{#1}}

\title{Measuring the CP property of  Higgs coupling to tau leptons in the VBF channel at the LHC}

\author[a,c]{Tao Han,}
\author[a]{Satyanarayan Mukhopadhyay,}
\author[b]{Biswarup Mukhopadhyaya}
\author[a,c]{and Yongcheng Wu}

\affiliation[a]{PITT-PACC, Department of Physics and Astronomy, University of Pittsburgh, PA 15260, USA}
\affiliation[b]{Regional Centre for Accelerator-based Particle Physics, Harish-Chandra Research Institute, HBNI, Chhatnag Road, Jhusi, Allahabad - 211019, India}
\affiliation[c]{Department of Physics, Tsinghua University and Collaborative Innovation Center of Quantum Matter, Beijing 100086, China}

\emailAdd{than@pitt.edu}
\emailAdd{satya@pitt.edu}
\emailAdd{biswarup@hri.res.in}
\emailAdd{wuyongcheng12@mails.tsinghua.edu.cn}

\abstract{We study the prospects of measuring the CP property of the Higgs ($h$) coupling to tau leptons  using the vector boson fusion (VBF) production mode at the high-luminosity LHC. Utilizing the previously proposed angle between the planes spanned by the momentum vectors of the $(\pi^+\pi^0)$ and $(\pi^- \pi^0)$  pairs originating in $\tau^\pm$ decays as the CP-odd observable, we perform a detailed Monte Carlo analysis, taking into account the relevant standard model backgrounds, as well as detector resolution effects. We find that 
excluding a pure CP-odd coupling hypothesis requires $\mathcal{O}(400 {~\rm fb}^{-1})$ luminosity at the $14$ TeV LHC, and 
values of the CP-mixing angle larger than about $25^\circ$ can be excluded at $95\%$ confidence level using $3 {~\rm ab}^{-1}$ data.
It is observed that the uncertainty in the angular resolution of the neutral pion momenta does not constitute a significant hurdle. Achieving a signal to background ratio ($S/B$) close to one, while keeping a high enough signal yield required to study the angular distributions selects out VBF as a promising mode to probe the CP nature of the $h\tau\tau$ coupling, with gluon fusion suffering from a low $S/B$, and the $W^\pm h/Zh$ mode (with leptonically decaying $W^\pm /Z$) having a much smaller signal rate.}

\begin{document} 
\maketitle
\flushbottom
\section{Introduction}
Measurement of the properties of the $125$ GeV scalar boson~\cite{Higgs} constitutes one of the most important experimental programs in the coming decades, at the ongoing $13$ TeV run of the LHC and its high-luminosity upgrade~\cite{HLLHC}, as well as at future planned lepton and higher-energy hadron colliders~\cite{ILC,CEPC,100TeV}. In addition to a precise determination of the coupling strengths of the Higgs boson ($h$) to standard model (SM) gauge bosons and fermions~\cite{Peskin,Plehn,Han}, future large statistics data sets would allow for the measurement of several differential distributions of interest, which are sensitive to the Lorentz structure and CP property of Higgs couplings~\cite{Dittmaier:2012vm, LHC_Higgs}. As is well known, extensions of the standard model, motivated by a range of arguments, predict observable deviations in the Higgs properties from their corresponding SM predictions~\cite{ILC,100TeV,LHC_Higgs2}. In particular, considerations on successful electroweak baryogenesis often require an extended Higgs sector, with potentially new sources of CP violation in Higgs couplings.

 ATLAS and CMS collaborations have utilized the $h \rightarrow Z Z^* \rightarrow (\ell^+ \ell^-) (\ell^{\prime +} \ell^{\prime-})$ decay mode (with $\ell , \ell^{\prime} = e, \mu$) to probe the CP nature of the Higgs coupling to $Z$-bosons, and using the $8$ TeV LHC data, a pure CP-odd hypothesis has been excluded at $99\%$ confidence level~\cite{CP_ZZ}. This, however, leaves open the possibility that $h$ is not an eigenstate of CP, but an admixture of CP-even and CP-odd states, and the present measurements do not constrain the amount of such a mixing significantly. CP violation in the $h \rightarrow Z Z^*$ decay process is not expected to be large as well, since the CP-even part of the amplitude originates from the $h Z^\mu Z_\mu$ term, which appears at dimension four, while the CP-odd coupling structure, $\epsilon_{\mu \nu \rho \sigma} h Z^{\mu \nu} Z^{\rho \sigma}$, can only be generated from dimension six SM gauge invariant operators. On the other hand, the Higgs coupling to SM fermions can include both CP-even and CP-odd Lorentz structures with similar strength, for example, via the mixing of CP-even and CP-odd scalars in a two-Higgs doublet model, and the CP-odd couplings can therefore be potentially large. Furthermore, the effective operators leading to the CP-odd Lorentz structures in the Higgs coupling to gauge bosons and to SM fermions (including different fermion generations) may have separate origin, thereby making the corresponding coupling strengths uncorrelated.

Among the third generation fermions, extensive studies have been performed to determine the LHC sensitivity of the structure of the Higgs coupling to top quarks~\cite{tth,Dorival}, and a realistic analysis including the effect of the relevant SM backgrounds shows that the high-luminosity (HL) LHC run with $3{~\rm ab}^{-1}$ of data can probe a CP-mixing angle larger than $\mathcal{O}(60^\circ)$ at $95\%$ C.L.~\cite{Dorival}. For the Higgs coupling to $b$-quarks, a measurement would be challenging, as no information of the $b$ polarization is retained in the angular distribution of the lightest $B$ meson's decay products, while only a fraction of the lightest baryon, $\Lambda_b$, is expected to retain the polarization information~\cite{Falk,Grossman}.

The fact that tau polarization information is retained in the angular distribution of its decay products~\cite{Hagiwara_Martin1,Bullock} has led to a number of studies on the prospects of measuring the CP property of the $h\tau^+\tau^-$ coupling at the LHC and at $e^+e^-$ colliders, utilizing both one and three prong $\tau$ decay modes~\cite{Nelson, Bower:2002zx, Desch:2003mw, Berge:2008wi, Berge:2008dr, Berge:2011ij, Berge:2013jra, Harnik:2013aja, Dolan:2014upa, Berge:2015nua, Askew:2015mda, Hagiwara:2016zqz}.  The existing phenomenological analyses have mostly focussed on the formulation of suitable observables that are sensitive to the CP nature of the $h\tau^+ \tau^-$ coupling, with the restriction that due to the presence of missing neutrinos, it is not possible to accurately determine the $\tau^\pm$ momentum vectors or the Higgs rest frame at the LHC. The one prong decay $\tau^\mp \rightarrow \rho^\mp \pbar{\nu_\tau}$, with $\rho^\mp \rightarrow \pi^\mp \pi_0$ is found to be promising in this regard~\cite{Bower:2002zx,Berge:2015nua}, as the angle between the decay planes spanned by $(\pi^+ \pi^0)$ and $(\pi^- \pi^0)$ can be utilized to define a CP-odd variable carrying the spin correlation information of the decaying taus. However, in order to properly access the viability of such a measurement at the LHC, it is important to study specific Higgs production modes, and determine to what extent such correlations can be extracted in the presence of large SM backgrounds.

The Higgs coupling to $\tau$ leptons has already been established with a combined significance of more than $5\sigma$~\cite{ATLAS_CMS} by the ATLAS~\cite{ATLAS_8TeV_VBF} and CMS~\cite{CMS_8TeV} collaborations, using the $7$ and $8$ TeV data from the LHC. The experimental analyses show the clear importance of the vector boson fusion (VBF) mode in driving the discovery of the $h \tau^+ \tau^-$ decay process. As is well known, VBF leads to a distinctive topology that can be used to enhance the signal to background ratio \cite{Rainwater}. In contrast, the leading Higgs production mode ($gg\to h\to \tau^+\tau^-$) suffers from formidable SM background \cite{Harnik:2013aja, Dolan:2014upa}, and the clean associated $W^\pm h/Zh$ production modes (with $h\rightarrow \tau^+ \tau^-$, and the vector boson decaying to leptonic final states)~\cite{ATLAS_ZH} suffer from very low rates of the signal itself as well as difficult backgrounds.

In this paper, we therefore focus on the vector boson fusion production of the Higgs boson in association with two forward tagging jets, and explore the prospects of probing the CP nature of the $h\tau^+\tau^-$ coupling using the $\tau^\mp \rightarrow  \pi^\mp \pi_0 \pbar{\nu_\tau}$ decay mode. We perform a complete Monte Carlo (MC) simulation of the signal and the dominant background processes, keeping the spin correlation in all steps of the decay chain, and including the effects of parton shower, hadronization and underlying events. We also carefully study the impact of the crucial detector resolution uncertainty in determining the momentum direction of the neutral pions.  As we shall show in the subsequent sections, VBF turns out to be a very promising mode for probing the $h\tau^+\tau^-$ coupling structure as well.

The rest of the paper is organized as follows. In Sec.~\ref{sec:analysis} we describe the parametrization of the effective interaction Lagrangian relevant for our study, the CP-odd observable defined in terms of the $\rho^\pm$ decay planes, and the details of our MC simulation of the signal and background processes. The kinematic selection criteria used to achieve an optimum signal to background ratio is described in the first part of Sec.~\ref{sec:results}, along with the signal and background rates for 14 TeV LHC. We then go on to discuss the CP-odd correlations, the impact of detector resolution uncertainties on them, the signal and background distributions after all cuts, and finally the projected measurement reach of the CP-mixing angle at the HL-LHC. We summarize our findings in Sec.~\ref{sec:summary}. The validation of our simulation framework against the ATLAS 8 TeV analysis of $h\rightarrow \tau^+ \tau^-$ in the VBF category is briefly discussed in the Appendix.

\section{Analysis Setup}
\label{sec:analysis}

\subsection{Effective interaction Lagrangian}
\label{ssec:eft}
We parametrize the effective $h\tau^+\tau^-$ interaction after electroweak symmetry breaking as follows:
\begin{equation}
\mathcal{L}_{h\tau\tau} = -\frac{y_\tau}{\sqrt{2}}h\bar{\tau}(\cos\Delta + i\gamma_5\sin\Delta)\tau.
\label{eq:L1}
\end{equation}
Here, $h$ is the observed $125$ GeV scalar mass eigenstate, $y_\tau$ is the Yukawa coupling of the tau lepton in the SM, $y_\tau = \sqrt{2} m_\tau/v$, with $m_\tau$ being the $\tau$ mass and $v \simeq 246$ GeV. With this parametrization, the branching fraction (BR) of $h \rightarrow \tau^+ \tau^-$ is fixed at its SM prediction, whose central value is $6.32\%$ for a $125$ GeV SM Higgs~\cite{Dittmaier:2011ti}.\footnote{In a more general parametrization of the $h\tau^+\tau^-$ effective interaction, the $h\rightarrow\tau^+\tau^-$ partial decay width can also be modified from its SM prediction.} Such an effective vertex can arise, for example, in a general two-Higgs doublet model, where the CP-even and CP-odd scalars can mix after electroweak symmetry breaking, thereby breaking CP symmetry. We shall refer to the angle $\Delta$ as the CP-mixing angle, which, in our convention, takes values in the range 
\begin{equation}
-\pi/2 < \Delta \leq \pi/2, 
\end{equation}
with $\Delta=0$ and $\Delta=\pi/2$ corresponding to pure CP-even and pure CP-odd couplings respectively. While discussing our results in later sections, in addition to $\Delta=0$ and $\Delta=\pi/2$, we shall also use $\Delta=\pi/4$ as a benchmark, which corresponds to a maximal mixing.

Although there are no direct constraints on $\Delta$ from any measurement so far, there can be indirect constraints from the upper bound on the electric dipole moment of electrons and neutrons. Such constraints on the CP-odd component of the $h\tau^+\tau^-$ coupling is rather weak at present due the smallness of the $\tau$ Yukawa coupling, and in fact they do not restrict the coupling range \cite{Brod:2013cka} considered in this study. Furthermore, the electric dipole moment constraints hold only under the further assumption that the first generation Yukawa couplings are the same as in the SM, making the direct collider measurement an important complementary probe.

As discussed earlier, for studying the CP property of the $h\tau^+\tau^-$ coupling, we shall focus on the one prong tau decay mode, $\tau^\mp \rightarrow \rho^\mp \pbar{\nu_\tau}$, which has the highest branching fraction of $25.49\%$~\cite{PDG} among hadronic one prong decays of tau leptons. The $\rho^\pm$ mesons subsequently decay to the fully reconstructible $\pi^\pm \pi^0$ final state, with nearly $100\%$ branching fraction. The effective interaction Lagrangian for the $\tau^\pm$ and $\rho^\pm$ decays can be parametrized as follows~\cite{Bullock}:
\begin{equation}
\label{eq:tau_decay}
\mathcal{L}_{\tau\rho\pi} = C_\tau (\rho^-_\mu \bar{\tau}\gamma^\mu P_L \nu_\tau + {\rm h.c.}) + C_\rho (\rho^-_\mu\pi_0\partial^\mu\pi^+ - \rho^-_\mu\pi^+\partial^\mu\pi_0 + {\rm h.c.}), 
\end{equation}
where $P_L$ denotes the chiral projection operator $P_L = (1-\gamma_5)/2$. In general, the effective couplings $C_\tau$ and $C_\rho$ contain form factors that depend on the energy scale of the process. However, for our purpose, it is adequate to set them as constants while generating the Monte Carlo events, and re-scale the total cross-section by the product of the corresponding $\tau^\pm$ and $\rho^\pm$ branching ratios.

\subsection{Observable sensitive to the CP structure of $h\tau^+\tau^-$ coupling}
\label{ssec:obs}
CP-odd observables that are sensitive to the CP-mixing angle $\Delta$ have been studied utilizing different $\tau^\pm$ decay modes~\cite{Nelson, Bower:2002zx, Desch:2003mw, Berge:2008wi, Berge:2008dr, Berge:2011ij, Berge:2013jra, Harnik:2013aja, Dolan:2014upa,Berge:2015nua, Askew:2015mda, Hagiwara:2016zqz}. For the $\tau^\mp \rightarrow \rho^\mp \pbar{\nu_\tau}$ decay mode, one particular advantage is that there are two visible particles from each decaying $\tau$, namely, charged ($\pi^\pm$) and neutral ($\pi_0$) pions. Even though the $\tau$ momentum and the decay plane itself cannot be reconstructed due to the missing neutrino, it has been shown that the $\pi^\pm$ and the two $\pi_0$'s retain the information of the $\tau^\pm$ polarization, and therefore the spin-correlation between the $\tau^+$ and $\tau^-$ can be utilized to define an observable that is sensitive to the CP-mixing angle $\Delta$~\cite{Bower:2002zx}. As discussed in detail in Refs.~\cite{Bower:2002zx,Berge:2015nua}, we can define the following CP-odd observable using the charged and neutral pion momenta as follows: 

\begin{equation}
 \Phi_{\rm CP} = \arccos(\hat{p}^{0-}_{\perp} \cdot \hat{p}^{0+}_{\perp}) \times \text{sgn}(\hat{p}^- \cdot (\hat{p}^{0-}_{\perp} \times \hat{p}^{0+}_{\perp}))
 \label{eq:phicp}
\end{equation}
Here, $\hat{p}^{\pm}$ and $\hat{p}^{0 \pm}$ denote, respectively, the unit vectors for the charged and neutral pion three momenta in the $\pi^+ \pi^-$ zero-momentum frame, and $\hat{p}^{0 +}$ ($\hat{p}^{0 -}$) further correspond to the neutral pion produced from $\rho^+$ ($\rho^-$) decay. The perpendicular components of the neutral pion momenta, $\hat{p}^{0 \pm}_\perp$ are defined to be transverse to the associated charged pion direction, $\hat{p}^{\pm}$. With this definition, $\Phi_{\rm CP}$ takes values in the range 
\begin{equation}
- \pi \leq \Phi_{\rm CP} \leq \pi. 
\end{equation}
We see that under CP transformation,  $\Phi_{\rm CP} \rightarrow  -\Phi_{\rm CP}$, thereby making it CP-odd, and therefore senstitive to CP violation in the $h\tau\tau$ coupling.

There is an additional subtlety in selecting the phase-space region of the charged and neutral pion pairs~\cite{Bower:2002zx,Berge:2015nua}. It turns out that the terms in the squared matrix element for the decay process $h \rightarrow \tau^+ \tau^- \rightarrow (\pi^+ \pi^0 {\bar\nu_\tau}) (\pi^- \pi^0 \nu_\tau)$ that are sensitive to both the CP-mixing parameter $\Delta$, as well as the $\tau^\pm$ spin vectors, are proportional to the product $Y=(E_{\pi^+}-E_{\pi^{0+}})\times(E_{\pi^-}-E_{\pi^{0-}})$. Here the energies are defined in the corresponding $\tau^\pm$ rest frames. The origin of this factor is the vertex factor of $(p^\mu_{\pi^\pm} -p^\mu_{\pi^0})$ appearing in $\rho^\pm$ decay. $Y$ is clearly not positive definite. Thus, if we integrate over charged and neutral pion momenta with both $Y>0$ and $Y<0$, the CP-mixing sensitive terms in the matrix element squared would average out to zero. Therefore, we need to separate the events into two different classes, one with $Y>0$ and the other with $Y<0$. The differential distributions of $\Phi_{\rm CP}$ in these two classes are related by a phase-shift of $\pi$. In order to keep both class of events at the same time in our analysis, while keeping the range $-\pi \leq \Phi_{\rm CP} \leq \pi$ unchanged, we adopt the following prescription for events with $Y<0$:

\begin{equation}
    \Phi_{\rm CP} \rightarrow 
\begin{cases}
    \Phi_{\rm CP}-\pi,& \text{if } Y < 0  \text{ and } 0\leq  \Phi_{\rm CP} \leq \pi\\
    \Phi_{\rm CP}+\pi,              & \text{if } Y < 0 \text{ and } -\pi\leq  \Phi_{\rm CP} \leq 0 .
\end{cases}
\end{equation}
Since the $\tau^\pm$ rest frames cannot be reconstructed at the LHC, we define the variable $Y$ using Lab frame energies instead, $Y=(E_{\pi^+}^L-E_{\pi^{0+}}^L)\times(E_{\pi^-}^L-E_{\pi^{0-}}^L)$. In the context of the 13 TeV LHC, it has been shown in Ref.~\cite{Berge:2015nua} that using Lab frame energies does not degrade the asymmetries significantly, as long as the $\rho^\pm$ mesons have transverse momenta larger than about $20$ GeV, which is almost always realized in the kinematic region of our interest.

\subsection{Monte Carlo simulation of signal and background processes}
\label{ssec:MC}
The signal process of our interest is Higgs production in association with at least two jets, $h+\geq 2$-jets, with the dominant background being $Z+\geq 2$-jets, at the $14$ TeV LHC. In order to keep the full spin-correlation between the Higgs decay products, we generate the parton level events with the $2\rightarrow 8$ matrix elements for the signal process (and similarly for the $Z+$jets background),  namely, 
\begin{equation}
p p \rightarrow h j j \rightarrow (\tau^+ \tau^-) j j\rightarrow (\rho^+ {\bar\nu_\tau}) (\rho^- \nu_\tau) j j  \rightarrow (\pi^+ \pi^0 {\bar\nu_\tau}) (\pi^- \pi^0  \nu_\tau) j j.
\end{equation}
The intermediate propagators, for the $h,\tau^\pm$ and $\rho^\pm$ are restricted to be on-shell, as the off-shell contributions to the signal rate is expected to be negligible. We generate the required model files for {\tt MadGraph5aMCNLO}~\cite{MG5} corresponding to the $\tau$ and $\rho$ decay Lagrangian in Eq.~\ref{eq:tau_decay} using {\tt FeynRules}~\cite{Feynrules}, and have also cross-checked our results with the implementation in {\tt TauDecayLibrary}~\cite{Taudecay}. 

The parton level events are generated using {\tt MadGraph5aMCNLO}, which are then passed onto {\tt PYTHIA6}~\cite{Pythia,Pythia8} for including the effects of parton shower, hadronization and underlying events. We should note here that the effect of parton shower and hadronization are only included for the tagging jets, since we already generate $\tau$ decay products at the colour-neutral hadron level. The factorization and renormalization scales have been kept at the default {\tt MadGraph5aMCNLO} event-by-event dynamic choice, while we have employed the default {\tt NN23LO1}~\cite{NNPDF,LHAPDF} PDF set. Jets are formed using the {\tt anti-kT}~\cite{antikt} algorithm with radius parameter $R=0.5$ using {\tt FastJet}~\cite{Fastjet}. For the background simulation of $Z+\geq 2$jets, although we have checked the effects of matrix-element (ME) parton shower (PS) merging on the kinematic observables, we stick to the $Z+2-$jets ME followed by PS, since otherwise in the high multiplicity final state, obtaining a large Monte Carlo (MC) statistics in the ME-PS merged event sample is beyond the scope of our computational resources. Finally, we use {\tt DELPHES3}~\cite{Delphes} for simulating the detector effects, with appropriate modifications to the default {\tt DELPHES3} options as described below. 

Tau jets are identified at the MC truth level within {\tt DELPHES3}, where a $\tau$ flavor identification is made in the event history. Apart from the tau jet reconstruction efficiency, a further tau identification efficiency of $60\%$ has been used in our simulations, with a light jet faking as a tau jet with $1\%$ efficiency. We further demand the presence of a single charged track with $p_T>1 $ GeV within the tau jet cone, as well as the presence of two additional photons (as a proxy for a $\pi_0$) as hits in the electromagnetic calorimeter (ECAL). Since the momentum direction of the neutral pion inside the tau jet (or equivalently the visible tau jet momentum direction) is sensitive to the angular resolution within the ECAL, in our {\tt DELPHES3} based detector simulation code, we pay particular attention in setting the granularity of the ECAL. 

The current LHC detectors constitute of different layers within the ECAL with different granularity in pseudo-rapidity ($\eta$) and azimuthal angle ($\phi$). For example, the first longitudinal layer in ATLAS ECAL has a high granularity in $\eta$ (between 0.003 and 0.006), crucial to provide discrimination between single photon showers and two photons from a $\pi_0$ decay, while the second  
layer has a granularity of $0.025 \times 0.025$ in $\eta \times \phi$~\cite{ATLAS_ECAL}. For simplicity, we have used the fixed granularity of $0.025 \times 0.025$ in the $\eta \times \phi$ plane. We shall demonstrate in the next section the variation in the reconstructed shape of $\Phi_{\rm CP}$ as the ECAL granularity is modified within a factor of two of the above average value. We note that in the realistic object reconstruction algorithms employed by the ATLAS and CMS collaborations, a photon or neutral pion direction in the ECAL is not determined by a single ECAL cell, but rather by a complex algorithm which utilizes the information of multiple ECAL cell hits from the electromagnetic shower of the incident photon(s), thereby obtaining a more precise directional information of the $\pi^0$ momentum.

\section{Results}
\label{sec:results}
\subsection{Kinematic selection of signal region : signal and background rates}
\label{ssec:cuts}
The kinematic selection criteria employed in the search for Higgs boson decaying to $\tau^+\tau^-$ in the VBF channel is well established~\cite{Rainwater}. Following the ATLAS search criteria for the hadronic tau decay mode, $h\rightarrow \tau_h \tau_h$, using the VBF category at the $8$ TeV LHC~\cite{ATLAS_8TeV_VBF}, we have optimized the kinematic cuts for the $14$ TeV centre of mass energy, in order to maximize the statistical significance of the search, at the same time ensuring a large signal rate. In the following, we show results with $\Delta=0$ for the signal rates, although in our parametrization (Eq.~\ref{eq:L1}), the total rate is independent of $\Delta$, and the kinematic cuts used are not sensitive to $\Delta$. The successive selection criteria applied are as follows:

\begin{table}
\centering
\resizebox{\textwidth}{!}{
\begin{tabular}{|c|c|c|c|c|c|}
\hline
Cuts & Higgs$+\geq2$-jets [fb] & $\epsilon_S$ & $Z+\geq2-$jets [fb] & $\epsilon_B$&$S/B$\\
\hline
Basic $+$ VBF cuts: & 4.56 & -- &  141.95 & --  & 1/31\\
2 $\tau$-jet identification and reconstruction & 0.69 & 0.15 & 6.75 & 0.05 &1/9.8 \\
$p_{T}^{j_1}>50$ GeV, $p_{T}^{j_2}>30$ GeV & 0.56 &0.81 & 5.60 & 0.83 &1/10\\
$\text{min}(\eta_{j_1},\eta_{j_2})<\eta_{\tau_1,\tau_2}<\text{max}(\eta_{j_1},\eta_{j_2})$ & 0.53 & 0.96 & 3.84& 0.69&1/7\\
$p_{T}^{h}>80$ GeV & 0.39 & 0.72  & 2.35& 0.61 &1/6 \\
$120 {~\rm GeV}<m_{\tau\tau} < 165$ GeV & 0.25 &0.64 & 0.28 & 0.12 & 1/1.1\\
\hline
\end{tabular}
}
\caption{\small \sl Signal and background cross-sections at the $14$ TeV LHC after different cuts on the kinematic observables; see text for details on the selection criteria. The efficiency of each cut on the signal ($\epsilon_S$) and background rate ($\epsilon_B$), along with the signal to background ratio ($S/B$) are also shown. We show results with $\Delta=0$ for the signal rates, although in our parametrization (Eq.~\ref{eq:L1}), the total rate is independent of $\Delta$.}
\label{tab:VBF14TeV}
\end{table}

\begin{enumerate}
\item {\bf Basic $+$ VBF cuts:} We require at least two light flavor jets (i.e., jets not tagged as $\tau-$ or $b-$jets), with transverse momentum $p_T^j>20$ GeV and pseudo-rapidity $|\eta_j|<5.0$. In addition, the jets should be separated in the pseudo-rapidity-azimuthal angle ($\phi$) plane by a distance $\Delta R_{j_1j_2} > 0.4$, with $\Delta R_{j_1j_2}=\sqrt{(\eta_{j_1}-\eta_{j_2})^2+(\phi_{j_1}-\phi_{j_2})^2}$. The first two light jets, ordered according to their $p_T$, are referred to as the tagging jets.

Since we consider only hadronic tau decays in this study, a veto on isolated leptons is imposed. Events with isolated electrons or muons with $p_{T_\ell}>20$ GeV and $|\eta_{\ell}|<2.5$ are rejected. 

The VBF topology cuts on the tagging jets are as follows: separation in pseudo-rapidity of $\Delta \eta_{j_1j_2} > 3.8$, the requirement that the tagging jets lie in opposite hemispheres of the detector, $\eta_{j_1} \times \eta_{j_2} <0$, and a large tagging jet invariant mass, $M_{j_1j_2}>500$ GeV.

With these cuts we obtain a signal cross-section of $4.6$ fb, with a $Z+$jets background rate of $142$ fb, see Table~\ref{tab:VBF14TeV}. The signal to background ratio ($S/B$) stands at $1/31$ at this stage, with the $\Delta \eta_{j_1j_2} > 3.8$ cut playing the most important role in bringing it to a reasonable level.

\item {\bf $\tau$-jet identification and reconstruction:} We require two reconstructed $\tau$-tagged jets with a basic selection cut of $p_T^{\tau_j}>20$ GeV and $|\eta_{\tau_j}|<2.5$, each containing one charged prong (track) of $p_T > 1$ GeV inside the jet cone. As discussed earlier, we further require the presence of two additional photons (as a proxy for a $\pi_0$) as hits in the electromagnetic calorimeter. With a $60\%$ efficiency for tau tagging, the two tau jet reconstruction efficiency turns out to be $\epsilon_S = 0.42$ for the signal process, while it is $\epsilon_B = 0.14$ for the $Z+$jets background, giving a factor of $3$ improvement in $S/B$. The primary reason for this difference stems from more non-central tau's produced in the $Z+\geq 2-$jets process, whereas for $h+\geq 2-$jets the tau jets are almost always centrally produced. The difference in the transverse momentum distribution of the tau jets plays a secondary role. 

\item {\bf Tagging jet $p_T$ cut:} Following the ATLAS $8$ TeV analysis, we demand the tagging jets to further pass a minimum transverse momentum cut, $p_{T}^{j_1}>50$ GeV, $p_{T}^{j_2}>30$ GeV. This cut does not help in increasing the $S/B$ ratio, as the signal events from weak boson fusion diagrams would have an average transverse momentum set by the mass scales of the $W$ and $Z$ bosons. However, it is useful for reducing fake backgrounds from multi-jet production, where two light jets fake as tau jets, and therefore, we have kept this cut the same as in the ATLAS $8$ study. 

\item {\bf Rapidity ordering between tagging jets and tau tagged jets:} This cut is also related to the VBF topology, where the two forward tagging jets lie at larger $|\eta|$ regions compared to the centrally produced tau's~\cite{Rainwater} (i.e., $\text{min}(\eta_{j_1},\eta_{j_2})<\eta_{\tau_1,\tau_2}<\text{max}(\eta_{j_1},\eta_{j_2})$), and leads to a further improvement of $1.4$ in $S/B$.

\item {\bf Higgs $p_T$ cut:} Here, $p_T$ of the Higgs stands for the magnitude of the vector sum of the transverse momenta of the visible Higgs decay products --- the two charged and neutral pions. We find that demanding $p_{T}^{h}>80$ GeV optimizes $S$ and $S/\sqrt{S+B}$, and as we can see leads to a modest improvement in $S/B$. The Higgs $p_T$ distribution is correlated with the minimum tagging jet $p_T$ requirement above.

\item {\bf Invariant mass of the tau jets:} The invariant mass of the tau jets is reconstructed using the collinear approximation for the invisible neutrino momenta~\cite{Rainwater}, which works reasonably well as the Higgs and the Z boson are sufficiently boosted due to the $p_T$ requirement above. In order not to lose too many signal events, we choose an asymmetric window for the invariant mass with a relaxed upper cut, $120 {~\rm GeV}<m_{\tau\tau} < 165$ GeV. In this invariant mass window, the tail of the signal $m_{\tau\tau}$ distribution is largely retained. This leads to a further improvement by a factor of $\sim 6$ in $S/B$, bringing it to the level of $1/1.1$. 
\end{enumerate}
While the $S/B$ ratio might be increased further by tightening the requirements on $\Delta \eta_{jj}$, $M_{jj}$ and $m_{\tau\tau}$, we abstain from applying stronger cuts, in order not to reduce the signal event yield further. After all the cuts above, the signal rate stands at $0.25$ fb, which corresponds to $750$ events after the accumulation of $3000 {~\rm fb}^{-1}$ data. As we shall consider the differential distribution of $\Phi_{\rm CP}$ as our observable to study the $h\tau\tau$ coupling structure, followed by a binned likelihood analysis, it is important to keep the statisitical fluctuation in each bin at a reasonably small level. With the above event yields, with $10$ bins in $\Phi_{\rm CP}$, the average statistical fluctuation in each bin for the combined signal and background events is found to be less than $8\%$.

\subsection{CP-odd correlations and measurement reach at HL-LHC}
\begin{figure}[htb!]
\centering 
\includegraphics[width=0.6\textwidth]{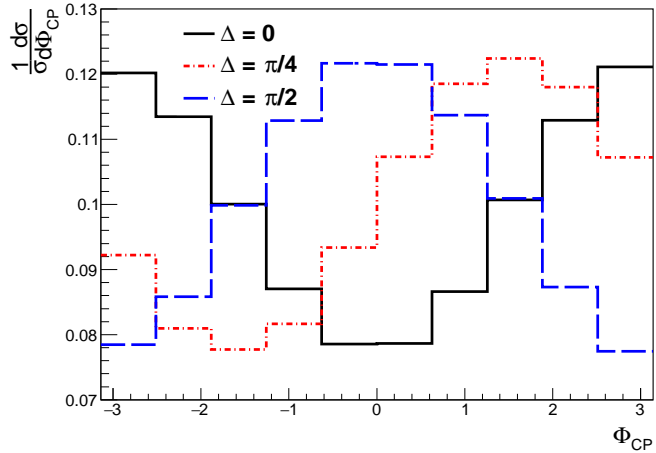}
\caption{\label{fig:parton}\small \sl Normalized differential distribution of $\Phi_{\rm CP}$ for the signal process of $h+\geq 2-$jets at the $14$ TeV LHC,  for three different values of the CP-mixing angle, $\Delta=0$ (black solid), $\pi/2$ (blue dashed) and $\pi/4$ (red dot-dashed). $\Phi_{\rm CP}$ is defined in terms of the momentum vectors of the charged and neutral pions in the $\pi^+ \pi^-$ zero momentum frame (Eq.~\ref{eq:phicp}). The distributions here are shown after the basic and VBF cuts, before folding in the detector effects..}
\end{figure}
Following the definition of $\Phi_{\rm CP}$ in Sec.~\ref{ssec:obs}, we show in Fig.~\ref{fig:parton} the normalized differential distribution of $\Phi_{\rm CP}$ for the signal process of $h+\geq 2-$jets at the $14$ TeV LHC,  for three different values of the CP-mixing angle, $\Delta=0$ (black solid), $\pi/2$ (blue dashed) and $\pi/4$ (red dot-dashed). We recall that as in Eq.~\ref{eq:phicp}, $\Phi_{\rm CP}$ is defined in terms of the momentum vectors of the charged and neutral pions in the $\pi^+ \pi^-$ zero momentum frame. In order to first understand the generic features of $\Phi_{\rm CP}$, the distributions in Fig.~\ref{fig:parton} are shown after the basic and VBF cuts discussed in the previous sub-section, before folding in the detector effects. The differential distribution of $\Phi_{\rm CP}$ can be described by a functional form given by \cite{Bower:2002zx,Berge:2015nua}
\begin{equation}
A - B \cos(\Phi_{CP} + 2 \Delta),
\end{equation}
 where $A$ corresponds to the total cross-section, and $B$ determines the relative magnitude of the asymmetry. Therefore, compared to the pure CP-even case  ($\Delta=0$), the $\Phi_{\rm CP}$ distribution (and correspondingly its minima and maxima) for a non-zero CP-mixing angle $\Delta$ is shifted by a phase of $2\Delta$, thereby giving us a clear discrimination of different CP-mixing angles.

\begin{figure}[htb!]
\centering 
\includegraphics[width=0.6\textwidth]{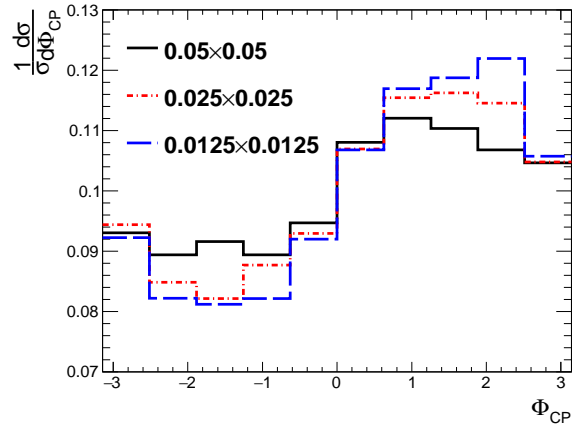}
\caption{\label{fig:detector}\small \sl Normalized differential distribution of $\Phi_{\rm CP}$ for the $h+\geq 2-$jets process at 14 TeV LHC, with $\Delta=\pi/4$, for three different choices of the electromagnetic calorimeter granularity. The blue dashed, red dot-dashed and the black solid histograms correspond to an ECAL granularity of $0.0125 \times 0.0125$, $0.025 \times 0.025$ and $0.05 \times 0.05$ respectively, in the $\eta \times \phi$ plane.}
\end{figure}
After folding in the detector effects as discussed in Sec.~\ref{ssec:MC}, we have checked that the charged pion transverse momentum resolution and the neutral pion energy resolution do not affect the distribution of $\Phi_{\rm CP}$ significantly. As emphasized in Sec.~\ref{ssec:MC}, the most important detector effect enters through the angular resolution of the momentum direction of the neutral pions, which in turn is determined by the granularity of the ECAL in the $\eta \times \phi$ plane. In order to understand the variation in the reconstructed shape of $\Phi_{\rm CP}$ as the ECAL granularity is varied within a factor of two of the average value adopted for our subsequent analysis, we show in Fig.~\ref{fig:detector}, the normalized differential distribution of $\Phi_{\rm CP}$ (for $\Delta=\pi/4$), with three different choices of the ECAL granularity. As before, the distributions are shown for the signal process of $h+\geq2-$jets at the 14 TeV LHC after the basic and VBF cuts. The blue dashed, red dot-dashed and the black solid histograms correspond to an ECAL granularity of $0.0125 \times 0.0125$, $0.025 \times 0.025$ and $0.05 \times 0.05$ respectively. As we can see from this figure, although we lose information of the asymmetry in $\Phi_{\rm CP}$ for a coarser granularity, even using a segmentation of $0.05 \times 0.05$ should not affect our results significantly. Most importantly, the positions of the minima and maxima remain unchanged. For all subsequent analyses, we shall use the average value of $0.025 \times 0.025$, which is the granularity for the second layer of the ATLAS ECAL~\cite{ATLAS_ECAL}.

\begin{figure}[htb!]
\centering 
\includegraphics[width=0.6\textwidth]{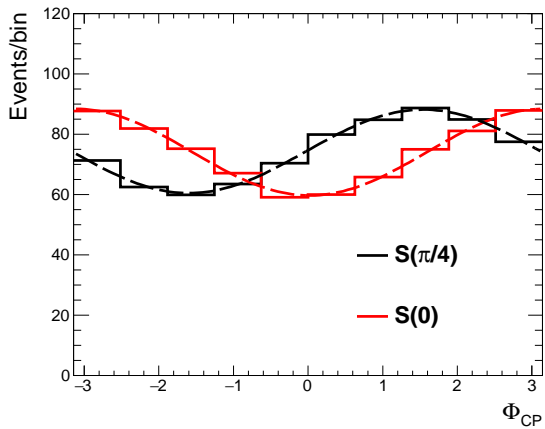}
\caption{\label{fig:cuts_S}\small \sl $\Phi_{\rm CP}$ distribution after all selection cuts as in Table~\ref{tab:VBF14TeV}, for the $h+\geq 2-$jets signal events, with the CP-mixing angle $\Delta=0$ (red) and $\Delta=\pi/4$ (black). The event numbers correspond to an integrated luminosity of $3000 {~\rm fb}^{-1}$ at the $14$ TeV LHC.}
\end{figure}
\begin{figure}[htb!]
\centering 
\includegraphics[width=0.6\textwidth]{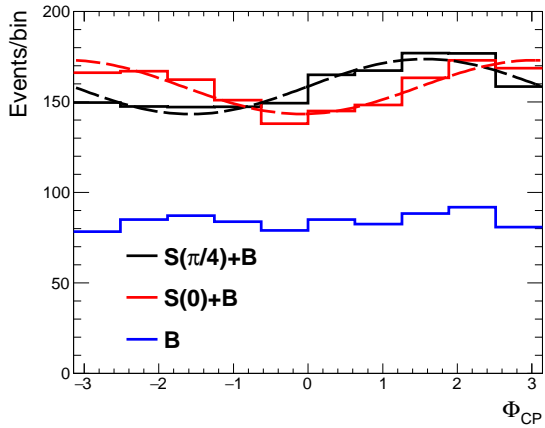}
\caption{\label{fig:cuts_S_B}\small \sl $\Phi_{\rm CP}$ distribution after all selection cuts as in Table~\ref{tab:VBF14TeV}, for the $Z+\geq 2-$jets background (blue), and the combined distribution of the $h+\geq 2-$jets signal and the background events, for the CP-mixing angle $\Delta=0$ (red) and $\Delta=\pi/4$ (black). The event numbers correspond to an integrated luminosity of $3000 {~\rm fb}^{-1}$ at the $14$ TeV LHC.}
\end{figure}

We are now in a position to show the $\Phi_{\rm CP}$ distributions for the signal, the $Z+\geq 2-$jets background, as well as the combined distribution of the signal and the background after all the kinematic selection cuts described in Sec.~\ref{ssec:cuts}. As seen in Table~\ref{tab:VBF14TeV}, after all cuts, the signal rate stands at $0.25$ fb, which corresponds to $750$ events after the accumulation of $3000 {~\rm fb}^{-1}$ data. The corresponding SM background cross-section is $0.28$ fb, which will lead to $840$ background events with the same luminosity, with an expected flat distribution in $\Phi_{\rm CP}$. In Fig.~\ref{fig:cuts_S} we compare the signal distributions for $\Delta=0$ (red histogram) and $\Delta=\pi/4$ (black histogram). It is encouraging to observe that, as expected, the kinematic selection criteria employed do not modify the shape of the signal distributions, and the two cases can be clearly distinguished with the above statistics. In Fig.~\ref{fig:cuts_S_B}, we show the $\Phi_{\rm CP}$ distribution for the $Z+\geq 2-$jets process after all cuts with the blue solid histogram. There are small fluctuations seen in the background distribution due to limited MC statistics (about $3000$ MC events remain after all cuts in our background sample). The red and black histograms in Fig.~\ref{fig:cuts_S_B} show the combined distribution of the signal and the background events for $\Delta=0$ and $\Delta=\pi/4$ respectively. Apart from the MC statistical fluctuation, we see that on adding the background events the difference in the positions of the minima and maxima between the two cases is still clearly distinguishable, which is promising. The most important reason for being able to achieve a positive discrimination is the fact that the VBF topology cuts and the tau jet pair invariant mass requirements help us achieve an $S/B$ ratio close to one. 

\begin{figure}[htb!]
\centering 
\includegraphics[width=0.6\textwidth]{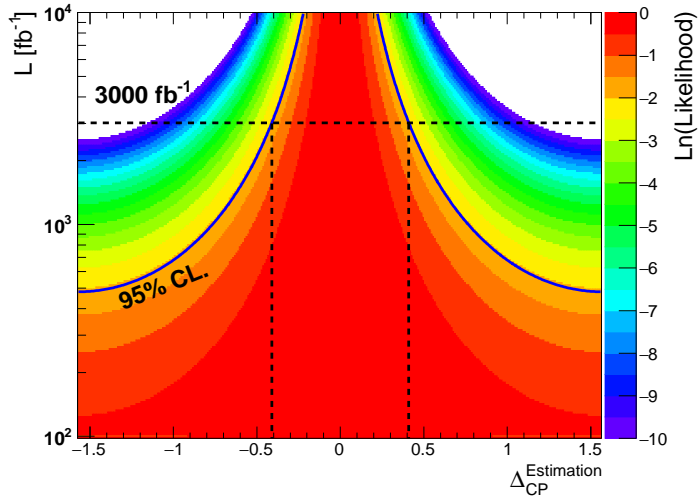}
\caption{\label{fig:reach}\small \sl Required integrated luminosity at the 14 TeV LHC, for achieving different values of the binned log-likelihood as a function of $\Delta$. The projected $95\%$ C.L. exclusion contours are shown by the solid blue curves.}
\end{figure}

Using the combined distributions for $\Phi_{\rm CP}$ of the signal and background events, as in Fig.~\ref{fig:cuts_S_B}, we perform a binned log-likelihood analysis where the probability of observing a certain number of events in each bin is modeled by the usual Poisson statistics~\footnote{To be specific, the log-likelihood is defined as 
$\mathcal{LL}=\sum_i\left[n_i\ln\left(\frac{n_i}{\nu_i}\right)+\nu_i-n_i\right]$, where, $n_i$ and $\nu_i$ are the observed and expected number of events in bin $i$, with the sum running over all bins.
}. The log likelihood is used to test the alternative hypothesis of a non-zero CP-mixing angle $\Delta$, against the null hypothesis of the CP even SM case with $\Delta=0$. In Fig.~\ref{fig:reach}, we show the required luminosity for achieving different values of the log-likelihood discriminant as a function of $\Delta$. The projected $95\%$ C.L. exclusion contours are shown by the solid blue curves. As we can see from this figure, excluding the pure CP-odd ($\Delta=\pi/2$) hypothesis at $95\%$ C.L. will require an integrated luminosity of $\mathcal{O}(400) {~\rm fb}^{-1}$ at the 14 TeV LHC, while CP-mixing angles in the range $[-90^\circ, -23^\circ]$ and $[23^\circ,90^\circ]$ can be probed with $3000 {~\rm fb}^{-1}$ of data at the HL-LHC. We note that the same method can also distinguish between the mixing angles $\Delta$ and $-\Delta$, where changing the sign of $\Delta$ induces a relative sign between the CP-even and the CP-odd terms in the effective Lagrangian. 

Although we did not present here a detailed study on the importance of the gluon fusion and vector boson associated production modes in probing the $h\tau^+ \tau^-$ coupling structure, a few remarks are in order. Utilizing the $\tau^\mp \rightarrow \rho^\mp \pbar{\nu_\tau}$ decay correlations (using a different CP sensitive observable~\cite{Harnik:2013aja}) in gluon fusion Higgs production, it was observed in Ref.~\cite{Askew:2015mda} that it is challenging to reduce the $Z$-induced background to the required level, and the maximally different pure CP-even ($\Delta=0$) and pure CP-odd ($\Delta=\pi/2$) hypotheses can be separated at the HL-LHC at most with $95\%$ C.L.  Even though $Zh$ and $W^\pm h$ production (with the vector boson decaying leptonically, and $h \rightarrow \tau^+ \tau^-$) are rather clean modes with a leptonic trigger, we find that the signal rate in these channels after necessary kinematic selection is significantly smaller than in VBF, and therefore studying the angular distributions with a reasonable accuracy is harder. However, even though the gluon fusion and the associated production modes are less sensitive than VBF, it is expected that a statistical combination of the analyses targeting different Higgs production and various tau decay modes would lead to an enhanced sensitivity on the CP-mixing angle $\Delta$.

\section{Summary}
\label{sec:summary}
In this paper, we revisit the prospects of measuring the CP structure of the $h \tau^+ \tau^-$ coupling at the high-luminosity upgrade of the 14 TeV LHC. We focus on the vector boson fusion production mode of the 125 GeV Higgs boson for this purpose, while previous detailed studies including the effects of SM backgrounds have mostly concentrated on the inclusive production of the Higgs with one jet, which is dominated by gluon fusion. As already demonstrated by the $8$ TeV ATLAS and CMS discovery analyses of the $h\rightarrow \tau^+\tau^-$ decay mode, the unique VBF signal topology offers an important handle to enhance the signal to background ratio. Enhancing the $S/B$ ratio is crucial to observe the angular correlation of the Higgs decay products above a large irreducible flat background distribution, mostly coming from $Z$ decays. In order to capture the spin correlation between the tau leptons, we work with the $\tau^\mp \rightarrow  \pi^\mp \pi_0 \pbar{\nu_\tau}$ decay mode, since, as shown in previous studies, the angle between the planes spanned by the $(\pi^+ \pi^0)$ and $(\pi^- \pi^0)$ momentum vectors can be utilized to define a CP-odd observable using only the visible Higgs decay products. 

For the signal process of $p p \rightarrow h j j \rightarrow (\tau^+ \tau^-) j j \rightarrow (\pi^+ \pi^0 {\bar\nu_\tau}) (\pi^- \pi^0  \nu_\tau) j j$, and similarly for the dominant SM background process of $Z+\geq 2-$jets, we have generated the matrix elements with full spin correlation, and subsequently included the effects of parton shower, hadronization, underlying events, and detector resolution. We pay special attention to the uncertainty coming from the angular resolution of the momentum direction of the neutral pions (and equivalently the visible tau jet momentum direction) due to the finite granularity of the electromagnetic calorimeter. It is encouraging to observe that using an ECAL granularity of $0.025 \times 0.025$ in the $\eta \times \phi$ plane (which is the average value in the second layer of the ATLAS ECAL) provides excellent resolution of the neutral pion momentum direction, and even a coarser granularity of $0.05 \times 0.05$ does not degrade the asymmetries significantly. In an actual experimental analysis, which employs a more complex algorithm to determine the position of a $\pi^0$ in the $\eta-\phi$ plane based on the shower profile of the photons from its decay, we expect that the angular correlation between the $(\pi^+ \pi^0)$ and $(\pi^- \pi^0)$ planes can be reconstructed with an even better accuracy. 

In order to enhance the signal to background ratio, we follow the ATLAS 8 TeV analysis in the search for $h \rightarrow \tau^+ \tau^-$ in the VBF category, and optimize the cuts on the kinematic variables for the 14 TeV centre of mass energy. Apart from the VBF selection cuts, the invariant mass of the tau pair, reconstructed using the collinear approximation for the missing neutrinos, provides the most useful handle to achieve an $S/B$ ratio close to one. The $h+\geq 2-$jets signal rate after all cuts is found to be $0.25$ fb, with a corresponding $Z+\geq 2-$jets background rate of $0.28$ fb, giving us $750$ and $840$ signal and background events respectively with $3000 {~\rm fb}^{-1}$ of data at the 14 TeV LHC, sufficient to study the angular correlations with a reasonable accuracy. The kinematic selection criteria do not affect the shape of the differential distribution for the CP-odd observable $\Phi_{\rm CP}$, and we find that with the above statistics, the combined distribution of $\Phi_{\rm CP}$ with both signal and background events shows a clear distinction between the SM hypothesis of CP-mixing angle $\Delta=0$, and the maximal CP violating case of $\Delta=\pi/4$, for example. 

We perform a binned log-likelihood analysis to determine the projected reach of the LHC in probing the CP-mixing angle. For $\Delta=\pi/2$, which corresponds to the pure CP-odd case and therefore leads to a maximal difference in the $\Phi_{\rm CP}$ distribution compared to the pure CP-even case, it is found that a $95\%$ C.L. exclusion can be achieved with $\mathcal{O}(400) {~\rm fb}^{-1}$ of data. With the accumulation of $3000 {~\rm fb}^{-1}$ integrated luminosity at the HL-LHC, CP-mixing angles larger than about $25^\circ$ can also be probed at the $95\%$ C.L. level, along with the corresponding negative range of $\Delta$. Thus measuring the CP structure of the $h\tau^+\tau^-$ coupling using the vector boson fusion mode holds sufficient promise and should be pursued at the HL-LHC. We expect that the experimental collaborations would carry out a further feasibility study of this measurement, including important effects such as that of pile-up, which were beyond the scope of our simulation framework.

\section*{Acknowledgments}
We thank Xin Chen for very helpful inputs. SM is thankful to Joseph Boudreau for discussions on the ATLAS ECAL, and James A. Mueller for help with questions on statistics.
TH and SM are supported in part by the U.S. Department of Energy under grant No. DE-FG02-95ER40896 and in part by the PITT PACC. B.M. is supported by the funding available from the Department of Atomic Energy, Government of India, for the Regional Centre for Accelerator based Particle Physics (RECAPP), Harish-Chandra Research Institute. B.M. would also like to thank the PITT PACC, University of Pittsburgh, for hospitality during the initial phase of the work.

\section*{Appendix: Validation of MC simulation with ATLAS 8 TeV analysis}
In order to validate our MC setup and detector simulation, we have followed the ATLAS cut-based analysis  in the VBF category for the Higgs decay to tau hadron pair channel, at the $8$ TeV LHC (with an integrated luminosity of $20.3 {~\rm fb}^{-1}$)~\cite{ATLAS_8TeV_VBF}. The selection cuts used are as summarized in Table 15 of Ref.~\cite{ATLAS_8TeV_VBF}. For the SM backgrounds, we have only included the dominant $Z+$jets process, and find a final event yield larger than the ATLAS expected total background estimates (which also includes fake backgrounds from multi-jet production, as well as other sub-dominant physics backgrounds). The cut-flow is presented in Table~\ref{tab:ATLAS_VBF_8TeV}, following Table 15 of Ref.~\cite{ATLAS_8TeV_VBF}.

\begin{table}[htb!]
\centering
\resizebox{\textwidth}{!}{
\begin{tabular}{|c|c|c|c|c|c|c|}
\hline
 & \multicolumn{2}{c|}{High $P_T$} & \multicolumn{2}{c|}{Low $P_T$ tight} & \multicolumn{2}{c|}{Low $P_T$ loose} \\
 \cline{2-7}
Cross section [fb] & $h+$jets & $Z+$jets & $h+$jets & $Z+$jets & $h+$jets &$Z+$jets   \\
 \hline
 Basic Selection Cuts & 350  & 1040760  & 350 & 1040760 & 350  & 1040760  \\
 $\met > 20$ GeV  & 197.45 & 298251  & 197.45  & 298251 & 197.45 & 298251  \\
  No Isolated Leptons & 156.55 & 276170  & 156.55 & 276170  & 156.55  & 276170  \\
  Two Tau-jets & 10.73 & 4393.09 & 10.73  & 4393.09 & 10.73 & 4393.09  \\
  At least two hadronic jets & 4.18 & 883.99  & 4.18 & 883.99 & 4.18 & 883.99  \\
  $P_T^{j_1}>50\text{ GeV}$, $P_T^{j_2}>30\text{ GeV}$ & 2.95 & 545.71  & 2.95 & 545.71 & 2.95 & 545.71  \\
  $|\Delta\eta_{jj}| > 2.6$ & 1.20  & 86.60  & 1.20 & 86.60 & 1.20 &  86.60 \\
  $m_{jj}>250$ GeV & 1.12 & 72.08 & 1.12  & 72.08 & 1.12 & 72.08  \\
\hline
  $0.8<\Delta R_{\tau_1\tau_2}<1.5$(high $P_T$)& 0.26 & 21.44 & -- & -- & -- & --  \\
  $0.8<\Delta R_{\tau_1\tau_2}<2.4$(low $P_T$) & -- & -- & 0.79 &  51.32 & 0.79 & 51.32 \\
 \hline
   $|\Delta\eta_{\tau_1\tau_2} |< 1.5 $& 0.26  & 21.44 & 0.71  &  48.46 & 0.71 &  48.46\\
 \hline
  $P_T^{H} > 140$ GeV (high $P_T$)& 0.25 & 14.45 & -- & -- & -- & --  \\
  $\Delta R_{\tau_1\tau_2}>1.5$ or $P_{T}^{H}<140$ GeV(low $P_T$) & -- & -- & 0.46 & 34.01 & 0.46 & 34.01 \\
 \hline
  $\met$  direction in $\phi$ & 0.25 & 14.37 & 0.46 & 33.56 & 0.46 & 33.56 \\
   ${\rm min}(\eta_{j_1},\eta_{j_2})<\eta_{\tau_1,\tau_2}<{\rm max}(\eta_{j_1},\eta_{j_2})$& 0.21  & 7.52 & 0.40 & 18.28 & 0.40 & 18.28 \\
\hline
  $m_{j_1,j_2} > -250 \cdot\vert\Delta\eta_{j_1,j_2}\vert + 1550$(tight) & -- & -- & 0.22 & 6.10  & -- & -- \\
  $m_{j_1,j_2} < -250 \cdot \vert\Delta\eta_{j_1,j_2}\vert + 1550$(loose) & -- & -- & -- & -- & 0.17 & 12.19\\
\hline
 Events with 20.3 fb$^{-1}$ & 4.29 & 152.70 & 4.56 & 123.73 &3.48 &247.43 \\
ATLAS Expected (central value) & 5.70  & 59 & 5.2 & 86 & 3.70  & 156.00 \\
\hline     
Ratio (Ours/ATLAS)&0.75    & 2.59   &   0.88      &  1.44    & 0.94	&1.59  \\

\hline
\end{tabular}
}
\caption{\small \sl The cut-flow for the signal process $h+$jets (with $h \rightarrow \tau_h \tau_h$) and the dominant SM background process of $Z+$jets (with $Z \rightarrow \tau_h \tau_h$) in the VBF category for three different signal regions as defined in Ref.~\cite{ATLAS_8TeV_VBF}. All cross-sections are shown in femtobarn units. Our results are shown at leading order for the 8 TeV LHC, whereas the ATLAS results include higher order corrections. For the SM backgrounds, we have only included the dominant process of $Z+$jets, while the ATLAS estimates shown also include fake backgrounds from multijet production, as well as other sub-dominant physics backgrounds.}
\label{tab:ATLAS_VBF_8TeV}
\end{table}

As we can see from Table~\ref{tab:ATLAS_VBF_8TeV}, even though our signal event numbers are close to the ATLAS ones (within a factor of $0.75 - 0.94$, which can be accounted for by higher order corrections not included in our analysis), our $Z+$jets background estimates are larger than the ATLAS total background numbers by a factor in the range of $1.4-2.6$. A detailed cut-flow analysis from ATLAS is not available for comparison, and we expect this difference to be stemming from our rather simplistic modelling of tau jet identification and reconstruction efficiency, as a function of their transverse momenta and pseudo-rapidity. However, since our background estimates are larger than the ATLAS estimates, while the signal numbers are very similar, our results can be considered as conservative. We note in passing that we have checked our inclusive leading order $Z+$jets production cross-section with the results reported in Ref.~\cite{Catani}, and find a very good agreement.



\begin{thebibliography}{99}
%
\bibitem{Higgs}
  G.~Aad {\it et al.} [ATLAS Collaboration],
  ``Observation of a new particle in the search for the Standard Model Higgs boson with the ATLAS detector at the LHC,''
  Phys.\ Lett.\ B {\bf 716} (2012) 1;
  S.~Chatrchyan {\it et al.} [CMS Collaboration],
  ``Observation of a new boson at a mass of 125 GeV with the CMS experiment at the LHC,''
  Phys.\ Lett.\ B {\bf 716} (2012) 30.
  
\bibitem{HLLHC}
ATLAS Phase-II Upgrade Scoping Document,
CERN-LHCC-2015-020,
{\tt https://cds.cern.ch/record/2055248}.
  
\bibitem{ILC}
  H.~Baer {\it et al.},
  ``The International Linear Collider Technical Design Report - Volume 2: Physics,''
  arXiv:1306.6352 [hep-ph].
  
\bibitem{CEPC}
CEPC Pre-CDR, 
{\tt http://cepc.ihep.ac.cn/preCDR/volume.html}.

\bibitem{100TeV}
  N.~Arkani-Hamed, T.~Han, M.~Mangano and L.~T.~Wang,
  ``Physics opportunities of a 100 TeV proton-proton collider,''
  Phys.\ Rept.\  {\bf 652} (2016) 1; 
%
  M.~L.~Mangano {\it et al.},
  ``Physics at a 100 TeV pp collider: Standard Model processes,''
  arXiv:1607.01831 [hep-ph].

\bibitem{Peskin}
  M.~E.~Peskin,
  ``Comparison of LHC and ILC Capabilities for Higgs Boson Coupling Measurements,''
  arXiv:1207.2516 [hep-ph].
 
 \bibitem{Plehn}
  M.~Klute, R.~Lafaye, T.~Plehn, M.~Rauch and D.~Zerwas,
  ``Measuring Higgs Couplings at a Linear Collider,''
  Europhys.\ Lett.\  {\bf 101} (2013) 51001.
  
 \bibitem{Han}
  T.~Han, Z.~Liu and J.~Sayre,
  ``Potential Precision on Higgs Couplings and Total Width at the ILC,''
  Phys.\ Rev.\ D {\bf 89} (2014) no.11,  113006.
  
\bibitem{Dittmaier:2012vm}
  S.~Dittmaier {\it et al.},
  ``Handbook of LHC Higgs Cross Sections: 2. Differential Distributions,''
  doi:10.5170/CERN-2012-002
  arXiv:1201.3084 [hep-ph].


  
\bibitem{LHC_Higgs}
  S.~Heinemeyer {\it et al.} [LHC Higgs Cross Section Working Group Collaboration],
  ``Handbook of LHC Higgs Cross Sections: 3. Higgs Properties,''
  doi:10.5170/CERN-2013-004
  arXiv:1307.1347 [hep-ph].  

\bibitem{LHC_Higgs2}
  D.~de Florian {\it et al.} [LHC Higgs Cross Section Working Group Collaboration],
  ``Handbook of LHC Higgs Cross Sections: 4. Deciphering the Nature of the Higgs Sector,''
  arXiv:1610.07922 [hep-ph].
    
 

\bibitem{CP_ZZ}
  G.~Aad {\it et al.} [ATLAS Collaboration],
  ``Evidence for the spin-0 nature of the Higgs boson using ATLAS data,''
  Phys.\ Lett.\ B {\bf 726} (2013) 120;
  S.~Chatrchyan {\it et al.} [CMS Collaboration],
  ``Study of the Mass and Spin-Parity of the Higgs Boson Candidate Via Its Decays to Z Boson Pairs,''
  Phys.\ Rev.\ Lett.\  {\bf 110} (2013) no.8,  081803;
  S.~Chatrchyan {\it et al.} [CMS Collaboration],
  ``Measurement of the properties of a Higgs boson in the four-lepton final state,''
  Phys.\ Rev.\ D {\bf 89} (2014) no.9,  092007.
  
\bibitem{tth}
	 N.~Mileo, K.~Kiers, A.~Szynkman, D.~Crane and E.~Gegner,arXiv:1603.03632 [hep-ph];
	 F.~Boudjema, R.~M.~Godbole, D.~Guadagnoli and K.~A.~Mohan, Phys.\ Rev.\ D {\bf 92}, no. 1, 015019 (2015);
	 J.~Ellis, D.~S.~Hwang, K.~Sakurai and M.~Takeuchi, JHEP {\bf 1404}, 004 (2014);
	 F.~Demartin, F.~Maltoni, K.~Mawatari, B.~Page and M.~Zaro, Eur.\ Phys.\ J.\ C {\bf 74}, no. 9, 3065 (2014);
	 S.~Biswas, R.~Frederix, E.~Gabrielli and B.~Mele, JHEP {\bf 1407}, 020 (2014).
  

\bibitem{Dorival}
  M.~R.~Buckley and D.~Goncalves,
  ``Boosting the Direct CP Measurement of the Higgs-Top Coupling,''
  Phys.\ Rev.\ Lett.\  {\bf 116} (2016) no.9,  091801.
  
\bibitem{Falk}
  A.~F.~Falk and M.~E.~Peskin,
  ``Production, decay, and polarization of excited heavy hadrons,''
  Phys.\ Rev.\ D {\bf 49} (1994) 3320.
  
\bibitem{Grossman}
  Y.~Grossman and I.~Nachshon,
  ``Hadronization, spin, and lifetimes,''
  JHEP {\bf 0807} (2008) 016;
  M.~Galanti, A.~Giammanco, Y.~Grossman, Y.~Kats, E.~Stamou and J.~Zupan,
  ``Heavy baryons as polarimeters at colliders,''
  JHEP {\bf 1511} (2015) 067.

\bibitem{Hagiwara_Martin1}
  K.~Hagiwara, A.~D.~Martin and D.~Zeppenfeld,
  ``Tau Polarization Measurements at LEP and SLC,''
  Phys.\ Lett.\ B {\bf 235} (1990) 198.
  
\bibitem{Bullock}
  B.~K.~Bullock, K.~Hagiwara and A.~D.~Martin,
  ``Tau polarization and its correlations as a probe of new physics,''
  Nucl.\ Phys.\ B {\bf 395} (1993) 499.

\bibitem{Nelson}
  J.~R.~Dell'Aquila and C.~A.~Nelson,
  ``{CP} Determination for New Spin Zero Mesons by the $\bar{\tau} \tau$ Decay Mode,''
  Nucl.\ Phys.\ B {\bf 320} (1989) 61.


\bibitem{Bower:2002zx}
  G.~R.~Bower, T.~Pierzchala, Z.~Was and M.~Worek,
  ``Measuring the Higgs boson's parity using tau $\rightarrow$ rho nu,''
  Phys.\ Lett.\ B {\bf 543} (2002) 227.

\bibitem{Desch:2003mw}
  K.~Desch, Z.~Was and M.~Worek,
  ``Measuring the Higgs boson parity at a linear collider using the tau impact parameter and tau $\rightarrow$ rho nu decay,''
  Eur.\ Phys.\ J.\ C {\bf 29} (2003) 491.

\bibitem{Berge:2008wi}
  S.~Berge, W.~Bernreuther and J.~Ziethe,
  ``Determining the CP parity of Higgs bosons at the LHC in their tau decay channels,''
  Phys.\ Rev.\ Lett.\  {\bf 100} (2008) 171605.

\bibitem{Berge:2008dr}
  S.~Berge and W.~Bernreuther,
  ``Determining the CP parity of Higgs bosons at the LHC in the tau to 1-prong decay channels,''
  Phys.\ Lett.\ B {\bf 671} (2009) 470.

\bibitem{Berge:2011ij}
  S.~Berge, W.~Bernreuther, B.~Niepelt and H.~Spiesberger,
  ``How to pin down the CP quantum numbers of a Higgs boson in its tau decays at the LHC,''
  Phys.\ Rev.\ D {\bf 84} (2011) 116003.


\bibitem{Berge:2013jra}
  S.~Berge, W.~Bernreuther and H.~Spiesberger,
  ``Higgs CP properties using the $\tau$ decay modes at the ILC,''
  Phys.\ Lett.\ B {\bf 727} (2013) 488.
  
\bibitem{Harnik:2013aja}
  R.~Harnik, A.~Martin, T.~Okui, R.~Primulando and F.~Yu,
  ``Measuring CP violation in $h \to \tau^+ \tau^-$ at colliders,''
  Phys.\ Rev.\ D {\bf 88} (2013) no.7,  076009.
  
\bibitem{Dolan:2014upa} 
  M.~J.~Dolan, P.~Harris, M.~Jankowiak and M.~Spannowsky,
  ``Constraining $CP$-violating Higgs Sectors at the LHC using gluon fusion,''
  Phys.\ Rev.\ D {\bf 90}, 073008 (2014).

\bibitem{Berge:2015nua}
  S.~Berge, W.~Bernreuther and S.~Kirchner,
  ``Prospects of constraining the Higgs boson's CP nature in the tau decay channel at the LHC,''
  Phys.\ Rev.\ D {\bf 92} (2015) 096012.
  
\bibitem{Askew:2015mda}
  A.~Askew, P.~Jaiswal, T.~Okui, H.~B.~Prosper and N.~Sato,
  ``Prospect for measuring the CP phase in the $h\tau\tau$ coupling at the LHC,''
  Phys.\ Rev.\ D {\bf 91} (2015) no.7,  075014.
  

\bibitem{Hagiwara:2016zqz}
  K.~Hagiwara, K.~Ma and S.~Mori,
  ``Probing CP violation in $h\to \tau^{-}\tau^{+}$ at the LHC,''
  arXiv:1609.00943 [hep-ph].
  
\bibitem{ATLAS_CMS}
  G.~Aad {\it et al.} [ATLAS and CMS Collaborations],
  ``Measurements of the Higgs boson production and decay rates and constraints on its couplings from a combined ATLAS and CMS analysis of the LHC pp collision data at $ \sqrt{s}=7 $ and 8 TeV,''
  JHEP {\bf 1608} (2016) 045.  

\bibitem{ATLAS_8TeV_VBF}
  G.~Aad {\it et al.} [ATLAS Collaboration],
  ``Evidence for the Higgs-boson Yukawa coupling to tau leptons with the ATLAS detector,''
  JHEP {\bf 1504} (2015) 117.
  
\bibitem{CMS_8TeV}
  S.~Chatrchyan {\it et al.} [CMS Collaboration],
  ``Evidence for the 125 GeV Higgs boson decaying to a pair of $\tau$ leptons,''
  JHEP {\bf 1405} (2014) 104.

\bibitem{Rainwater}
  D.~L.~Rainwater, D.~Zeppenfeld and K.~Hagiwara,
  ``Searching for $H\to\tau^+\tau^-$ in weak boson fusion at the CERN LHC,''
  Phys.\ Rev.\ D {\bf 59} (1998) 014037.
 
\bibitem{ATLAS_ZH} 
  G.~Aad {\it et al.} [ATLAS Collaboration],
  ``Search for the Standard Model Higgs boson produced in association with a vector boson and decaying into a tau pair in $pp$ collisions at $\sqrt{s} = 8$ TeV with the ATLAS detector,''
  Phys.\ Rev.\ D {\bf 93} (2016) no.9,  092005.


\bibitem{Dittmaier:2011ti}
  S.~Dittmaier {\it et al.} [LHC Higgs Cross Section Working Group Collaboration],
  ``Handbook of LHC Higgs Cross Sections: 1. Inclusive Observables,''
  doi:10.5170/CERN-2011-002
  arXiv:1101.0593 [hep-ph].

 

  
\bibitem{Brod:2013cka}
  J.~Brod, U.~Haisch and J.~Zupan,
  ``Constraints on CP-violating Higgs couplings to the third generation,''
  JHEP {\bf 1311} (2013) 180.

\bibitem{PDG}
  C.~Patrignani {\it et al.} [Particle Data Group Collaboration],
  ``Review of Particle Physics,''
  Chin.\ Phys.\ C {\bf 40} (2016) no.10,  100001.

  

%
%
\bibitem{MG5}
  J.~Alwall, M.~Herquet, F.~Maltoni, O.~Mattelaer and T.~Stelzer,
  ``MadGraph 5 : Going Beyond,''
  JHEP {\bf 1106}, 128 (2011);
  J.~Alwall {\it et al.},
  ``The automated computation of tree-level and next-to-leading order differential cross sections, and their matching to parton shower simulations,''
  JHEP {\bf 1407} (2014) 079.

\bibitem{Feynrules}
  N.~D.~Christensen and C.~Duhr,
  ``FeynRules - Feynman rules made easy,''
  Comput.\ Phys.\ Commun.\  {\bf 180} (2009) 1614;
  A.~Alloul, N.~D.~Christensen, C.~Degrande, C.~Duhr and B.~Fuks,
  ``FeynRules  2.0 - A complete toolbox for tree-level phenomenology,''
  Comput.\ Phys.\ Commun.\  {\bf 185} (2014) 2250.

\bibitem{Taudecay}
  K.~Hagiwara, T.~Li, K.~Mawatari and J.~Nakamura,
  ``TauDecay: a library to simulate polarized tau decays via FeynRules and MadGraph5,''
  Eur.\ Phys.\ J.\ C {\bf 73} (2013) 2489.

\bibitem{Pythia}
  T.~Sjostrand, S.~Mrenna and P.~Z.~Skands,
  ``PYTHIA 6.4 Physics and Manual,''
  JHEP {\bf 0605}, 026 (2006).
  
 \bibitem{Pythia8}  
  T.~Sjostrand, S.~Mrenna and P.~Z.~Skands,
  ``A Brief Introduction to PYTHIA 8.1,''
  Comput.\ Phys.\ Commun.\  {\bf 178} (2008) 852;
  T.~Sjostrand, S.~Ask, J.~R.~Christiansen, R.~Corke, N.~Desai, P.~Ilten, S.~Mrenna and S.~Prestel {\it et al.},
  ``An Introduction to PYTHIA 8.2,''
  Comput.\ Phys.\ Commun.\  {\bf 191} (2015) 159.
  
  
%
\bibitem{NNPDF}
  R.~D.~Ball {\it et al.},
  ``Parton distributions with LHC data,''
  Nucl.\ Phys.\ B {\bf 867} (2013) 244;
  R.~D.~Ball {\it et al.} [NNPDF Collaboration],
  ``Parton distributions for the LHC Run II,''
  JHEP {\bf 1504} (2015) 040.




\bibitem{LHAPDF}
  M.~R.~Whalley, D.~Bourilkov and R.~C.~Group,
  ``The Les Houches accord PDFs (LHAPDF) and LHAGLUE,''
  hep-ph/0508110.


 \bibitem{antikt}
  M.~Cacciari, G.~P.~Salam and G.~Soyez,
  ``The Anti-k(t) jet clustering algorithm,''
  JHEP {\bf 0804} (2008) 063.
%
%


\bibitem{Fastjet}
  M.~Cacciari, G.~P.~Salam and G.~Soyez,
  ``FastJet user manual,''
  Eur.\ Phys.\ J.\ C {\bf 72} (2012) 1896;
  M.~Cacciari and G.~P.~Salam,
  ``Dispelling the $N^{3}$ myth for the $k_t$ jet-finder,''
  Phys.\ Lett.\ B\ {\bf 641} (2006) 57.
  
    \bibitem{Delphes}
  S.~Ovyn, X.~Rouby and V.~Lemaitre,
  ``DELPHES, a framework for fast simulation of a generic collider experiment,''
  arXiv:0903.2225 [hep-ph];
  J.~de Favereau {\it et al.} [DELPHES 3 Collaboration],
  ``DELPHES 3, A modular framework for fast simulation of a generic collider experiment,''
  JHEP {\bf 1402} (2014) 057.


  
\bibitem{ATLAS_ECAL}
  G.~Aad {\it et al.} [ATLAS Collaboration],
  ``Search for the Standard Model Higgs boson in the two photon decay channel with the ATLAS detector at the LHC,''
  Phys.\ Lett.\ B {\bf 705} (2011) 452.
  
   \bibitem{Catani}
  S.~Catani, L.~Cieri, G.~Ferrera, D.~de Florian and M.~Grazzini,
  ``Vector boson production at hadron colliders: a fully exclusive QCD calculation at NNLO,''
  Phys.\ Rev.\ Lett.\  {\bf 103} (2009) 082001.
  



%
\end{thebibliography}
 \end{document}